\documentclass[prd,floatfix,twocolumn,amsmath,amssymb,floatfix]{revtex4}
\usepackage{graphicx,color,dcolumn,booktabs,bm}
\usepackage{subfigure}
\usepackage{longtable,lscape}
\usepackage{amssymb}
\usepackage{indentfirst}
\usepackage{epsfig}
\usepackage{feynmf}   %{feynmp}
\usepackage{epstopdf}   %{feynmp}
\usepackage{slashed}  %for Feynman symbols
\usepackage{cases}
\usepackage[pdfpages]{xcolor}
\definecolor{maroon}{RGB}{139,25,150}%burada 0-255 arasi her biri icin numara vererek renk elde et
\usepackage{multirow}
\usepackage{float}
\usepackage{graphicx,color,dcolumn,booktabs,bm}
\usepackage[colorlinks, citecolor=blue,anchorcolor=red,menucolor=red, linkcolor=red,filecolor=red,runcolor=red,urlcolor=blue,frenchlinks=red, urlcolor=blue]{hyperref}

\begin{document}
%=====================================================================================
%=====================================================================================
\title{\color{maroon}{ Properties of $Z_c(3900)$ tetraquark  in a cold nuclear matter }}
%=====================================================================================
%=====================================================================================
%
\author{K. Azizi$^{1,2}$}
%\email{kazizi@dogus.edu.tr}
\author{N. Er$^3$}
%\email{nuray@ibu.edu.tr}
\affiliation{
$^1$Department of Physics, University of Tehran, North Karegar Avenue, Tehran 14395-547, Iran \\
$^2$Department of Physics, Dogus University, Acibadem-Kadikoy, 34722 
Istanbul, Turkey \\
$^3$Department of Physics, Abant \.{I}zzet Baysal University,
G\"olk\"oy Kamp\"us\"u, 14980 Bolu, Turkey
}
%
%%%%%%%%%%%%%%%%%%%%%%%%%%%%%%%%%%%%%%%%%%%%%%%%%%%%%%%%%%%%%%%%%%%%%%%%%%%%%%%

\begin{abstract}
The study of medium effects on properties of particles  embedded in nuclear matter is of great importance for  understanding the nature and  internal quark-gluon organization as well as  exact determination of the  quantum numbers, especially of the exotic states. In this context, we study the physical properties of one of the famous charmonium-like states,  $Z_c(3900)$, in a cold dense matter. 	We  investigate the possible shifts in the mass and  current-meson coupling  of the $Z_c(3900)$ state due to the dense medium at saturation density, $ \rho^{sat}  $, by means of the in-medium sum rules. We also estimate  the vector self-energy of this state at saturation nuclear matter density.  We discuss the behavior of the spectroscopic parameters of this state with respect to the density up to a high density corresponding to the core of neutron stars, $\rho\approx 5\rho^{sat}$.  Both the mass and current-coupling of this state show nonlinear behavior and decrease with respect to the density of the medium: the mass reaches   roughly $30\%$  of its vacuum value at $ \rho=5\rho^{sat} $, while the current-coupling approaches  zero at $ \rho\approx2.1\rho^{sat} $, when the central values of the auxiliary and other input parameters are used. 
\end{abstract}

%%%%%%%%%%%%%%%%%%%%%%%%%%%%%%%%%%%%%%%%%%%%%%%%%%%%%%%%%%%%%%%%%%%%%%%%%%%%%%

%\pacs{21.65.-f,  11.55.Hx} 
% nuclear matter, sum rules

%\pacs{21.65.-f, 14.20.Mr, 14.20.Lq, 11.55.Hx} 
% nuclear matter, bottom baryons, changed baryons, sum rules
\maketitle

%%%%%%%%%%%%%%%%%%%%%%%%%%%%%%%%%%%%%%%%%%%%%%%%%%%%%%%%%%%%%%%%%%%%%%%%%%%%%%

\pagenumbering{arabic}

%%%%%%%%%%%%%%%%%%%%%%%%%%%%%%%%%%%%%%%%%%%%%%%%%%%%%%%%%%%%%%%%%%%%%%%%%%%%%%
%%%%%%%%%%%%%%%%%%%%%%%%%%%%%%%%%%%%%%%%%%%%%%%%%%%%%%%%%%%%%%%%%%%%%%%%%%%%%%

\section{Introduction}
The standard hadrons are divided into $ q\bar{q} $ and $ qqq/ \bar{q}\bar{q}\bar{q}$ systems. Neither the quark model nor the QCD  excludes the existence of the structures out of these configurations. Hence, search for exotic states is inevitable: now we have many exotic states observed in the experiment. We have also made good progress in the determination of different aspects of these states in theory. Most of the discovered exotic states are tetraquarks of the $ XYZ $ family. The term $ XYZ $ comes from the generalization of states $X(3872)$, $Y(4260)$  and $Z_c(3900)$. The $ XYZ $ states are  the charmonium/bottomonium-like resonances, which, because of their mass, can not be placed in the charmonium/bottomonium picture:   These  resonances,  mainly with $Q\bar{Q}q\bar{q}$  quark content, have different properties  than the standard excited quarkonia states.  In the last decade, many $XYZ$ states have been observed by the Belle, BESIII, BaBaR, LHCb, CMS, D0, CDF, and CLEO-c collaborations \cite{PhysRevD.98.030001}, and their masses,  widths and quantum numbers  $J^{PC}$ have been predicted. Detailed analysis of the experimental status of these states and various theoretical models can be found in numerous new review articles \cite{LIU2019237,RevModPhys.90.015004,ALI2017123,RevModPhys.90.015003}.

In the first trials, the new charmonium-like resonances, discovered in the above experiments, were evaluated as the excited states of ordinary $c\bar{c}$ charmonium states. However, the obtained data  showed that some resonances do not conform to standard spectroscopy and thus some new non-conventional models have been developed. The new models  differ from each other in terms of their components and strong interaction mechanisms. There are plenty of studies for the physical picture interpretation of the charmonium-like states using the new models:  for instance,  QCD tetraquark \cite{PhysRevD.100.014033,Luo2017}, weakly bound hadronic molecules \cite{Baru2017,CLOSE2004119,PhysRevD.76.094028,Coito2013,PhysRevD.77.034003,Cui_2014,PhysRevLett.111.132003}, charmonium hybrids \cite{ESPOSITO2016292,PhysRevLett.54.869,Liu2012,Li2013,PhysRevD.87.116004}, threshold cups \cite{PhysRevD.91.094025,Swanson:2015bsa,PAKHLOV2015183,PhysRevD.91.051504} and hadro-charmonium \cite{DUBYNSKIY2008344,Li:2013ssa}. Understanding the non-perturbative behaviour of QCD and the strong interaction dynamics that cause the production and structure of these non-conventional states is  very important issue for today's experimental and theoretical studies. 

As  candidates of tetraquark states, $Z^{\pm}_c(3900)$ were reported simultaneously by the BESIII Collaboration  \cite{PhysRevLett.110.252001} using  $e^+e^-$ annihilation at the vector resonance $Y(4260)$ and by the Belle Collaboration \cite{PhysRevLett.110.252002}using  the same process but at or near the $\Upsilon(nS)$ ($n = 1,\ 2,\ ...,\ 5$) resonances. For the full history see \cite{PhysRevLett.95.142001,PhysRevD.74.091104,PhysRevLett.99.182004}. They were confirmed by the CLEO-c Collaboration using 586 pb$^{-1}$ of $e^+e^-$ annihilation data taken at the CESR collider at $\sqrt{s}=4170$ MeV, the peak of the charmonium resonance $\psi(4160)$. They also reported evidence for $Z_c^0(3900)$ which is the neutral member of this isospin triplet  \cite{XIAO2013366}. In Ref. \cite{PhysRevLett.112.022001}  $e^+e^- \rightarrow \pi^{\pm}(D\bar{D}^*)^{\mp}$ at $\sqrt{s}=4.26$ GeV using a 525 pb$^{-1}$ data collected at the BEPCII storage ring, the determined  pole mass $M_{\textrm{pole}}=(3883.9\pm 1.5$(stat)$\pm4.2$ (syst)) MeV/$c^2$ and pole width  $\Gamma_{\textrm{pole}}=(24.8\pm 3.3$(stat)$\pm11.0$(syst)) MeV were reported  with significance of $2\sigma$ and $1\sigma$, respectively. It was referred as $Z_c(3885)$.  BESIII also observed  a new neutral state $Z_c(3900)^{0}$  in a process $e^+e^-\rightarrow \pi^0\pi^0 J/\psi$ with a significance of  $10.4\sigma$. The measured mass and width were $(3894.8\pm 2.3$(stat)$\pm3.2$(syst)) MeV/$c^2$ and $\Gamma=(29.6\pm 8.2$(stat)$\pm8.2$ (syst)) MeV, respectively  \cite{PhysRevLett.115.112003}.  In Ref. \cite{PhysRevD.92.092006}, after the full construction of the $D$ meson pair and the bachelor $\pi^{\pm}$ in the final state, the existence  of the charged structure $Z^{\pm}_c(3885)$ was confirmed  in the $(D\bar{D}^*)^{\mp}$ system and its pole mass and width were measured as $M_{\textrm{pole}}=(3881.7\pm 1.6$(stat)$\pm1.6$(syst)) MeV/$c^2$ and pole width  $\Gamma_{\textrm{pole}}=(26.6\pm 2.0$(stat)$\pm 2.1$(syst)) MeV, respectively. In the processes $e^+e^- \rightarrow D^+D^{*-} \pi^{0} + c.c.$ and $e^+e^- \rightarrow D^0\bar{D}^{*0} \pi^{0} + c.c.$ at $\sqrt{s}=4.226$ and $4.257$ GeV the neutral structure $Z_c(3885)^0$ was observed with the pole mass $(3885.7^{+4.3}_{-5.7}$(stat)$\pm8.4$ (syst)) MeV/$c^2$ and pole width $(35^{+11}_{-12}$(stat)$\pm15$ (syst)) MeV \cite{PhysRevLett.115.222002}. In a very recent study by the  D0 Collaboration,  the authors presented evidence for the $Z^{\pm}_c(3900)$ state decaying to $J/\psi\pi^{\pm}$ in semi-inclusive weak decays of $b-$flavored hadrons \cite{PhysRevD.98.052010}.

On theoretical side, for investigation of $Z_c(3900)$ resonance, a plenty of different models and approaches are used, some of them are mentioned here as examples. In a recent study \cite{Liu:2019gmh}, using the three- channel Ross-Shaw theory the authors have obtained constraint conditions that need to be satisfied by various parameters of the theory in order to have a narrow resonance close to the threshold of the third channel, it is relevant to the  structure.  Using the QCD sum rule method,  the same state was considered as a compact tetraquark state of diquark-antidiquark configuration in \cite{Kisslinger2015,PhysRevD.93.074002,PhysRevD.96.034026} and a hadronic molecule in \cite{PhysRevD.92.054002,PhysRevD.88.014030}. In these studies, many parameters related to the $Z_c(3900)$ state were calculated.  Its mass was already calculated within the framework of a non-relativistic  quark model in \cite{Patel2014}, as well.
 
 Despite a lot of  theoretical and experimental efforts, unfortunately, the nature and internal structures of most of the exotic states including  $Z_c(3900)$ are not exactly clear. Hence, investigation of their properties in a dense/hot medium can play an important role. Experiments like PANDA will provide a  possibility  to explore these states in a dense medium. These studies will help us better understand the quark-gluon organization of the exotic states as well as their interactions with the particles existing in the medium.  In our previous study,  Ref. \cite{AZIZI2018151},  we investigated the $X(3872)$ state by applying a diquark-antidiquark type current in the frame work of in-medium QCD sum rules. We  calculated the  mass, current-meson coupling and also the vector self-energy of this state  and found that these parameters strongly depend on the density of the medium. In the present study, we investigate the effects of a dense medium on the parameters of the $Z_c(3900)$ state and look for the behavior of the  mass,  current-meson coupling and vector self-energy of this charmonium-like state  considering it as a compact  tetraquark state.

The study is organized as follows. In Sec. II, we derive  the spectral densities associated with the state $Z_c(3900)$ by applying the two-point sum rule technique and obtain the QCD sum rules for the mass, current-meson coupling and vector self-energy using the obtained spectral densities. In section III, after fixing the auxiliary parameters using the standard prescriptions of the method,  the numerical analysis of the physical observables both in vacuum and cold nuclear matter is performed.  Section IV is devoted to discussion and  concluding remarks.  We collect some lengthy expressions obtained from the calculations in the Appendix.

%%%%%%%%%%%%%%%%%%%%%%%%%%%%%%%%%%%%%%%%%%%%%%%%%%%%%%%%%%%%%%%%%%%%%%%%%%%%%%
%%%%%%%%%%%%%%%%%%%%%%%%%%%%%%%%%%%%%%%%%%%%%%%%%%%%%%%%%%%%%%%%%%%%%%%%%%%%%%

\section{In-medim mass and current couplings of  $Z_c(3900)$}

Hadrons are formed as a  result of some non-perturbative effects at low energies very far from  the asymptotic region of QCD. Hence, for investigation of their properties, some non-perturbative methods are needed.  QCD sum rule  appears as a reliable, powerful and predictive approach in this respect. In this approach, the hadrons are represented by interpolating currents, written considering the quark content and all the quantum numbers of hadrons.  From the theoretical  studies in vacuum and the experimental data, the quantum numbers $J^{PC}=1^{+-}$  have been assigned to $Z_c (3900)$. The comparison of the theoretical predictions on some parameters of this state with the experimental data leads us consider a compact tetraquark of a  diquark and an antidiquark structure for this state \cite{PhysRevD.93.074002,PhysRevD.96.034026}. Thus, the interpolating current representing the  $\bar{c}cu\bar{d}$ quark content and $J^{PC}=1^{+-}$ quantum numbers can be written as
\begin{eqnarray}\label{ }
J_{\mu}(x)&=&\frac{i\epsilon_{abc}\epsilon_{dec}}{\sqrt{2}} \Bigg \{ \Big[u_{a}^T(x) C \gamma_5 c_b(x)\Big] \Big[\bar{d}_{d}(x) \gamma_{\mu} C\bar{c}^T_e(x) \Big] \nonumber \\
&-& \Big[u_{a}^T(x) C \gamma_{\mu} c_b(x)\Big] \Big[\bar{d}_{d}(x) \gamma_5 C\bar{c}^T_e(x) \Big] \Bigg \},
\end{eqnarray}
where $\epsilon_{abc}$ and $\epsilon_{dec}$ are anti-symmetric Levi-Civita symbols in three dimensions with color indices $a, b, c, d$ and $e$. The letter $T$ represents a transpose in Dirac space, $\gamma_5$ and $\gamma_{\mu}$ are Dirac matrices and $C$ is the charge conjugation operator.

In the framework of QCD sum rule, we aim to obtain  the sum rules for the  mass, current coupling and vector self-energy of  the exotic state $Z_c$ in cold nuclear matter. In the generalization of the vacuum QCD sum rules to finite density, we start with the same  correlation function of interpolating currents as the vacuum with a difference  that, in this case, the time ordering product of the interpolating currents is sandwiched between the ground states of a finite density medium instead of vacuum \cite{COHEN1995221}. So, we start with the following in-medium  two point correlation function as the building block of the method:
\begin{equation}\label{corre1}
\Pi_{\mu\nu}(p)=i\int{d^4 xe^{ip\cdot x}\langle\psi_0|\mathcal{T}[J_{\mu}(x)J^{\dagger}_{\nu}(0)]|\psi_0\rangle},
\end{equation}
where $p$ is the four momentum of the $Z_c$ state, $\mathcal{T}$ is the time ordering operator,  and $|\psi_0\rangle$ is the parity and time-reversal symmetric ground state of  nuclear matter. 

In cold nuclear matter, we formulate the sum rules for  the modified mass, $m_{Z_c}^*$, and current coupling constant, $f_{Z_c}^*$, of the ground state $Z_c$. For this purpose and in accordance with the philosophy of the method, the correlation function is calculated in two different perspectives: the phenomenological ($Phe$)  and operator product expansion  ($OPE$) windows. The results of these two representations are then matched under some conditions like the quark-hadron duality assumption to relate the hadronic parameters to fundamental QCD parameters. We deal with the ground state, hence, we should apply some transformations like the Borel and continuum subtraction to enhance the ground state contribution and suppress the contributions coming from higher states and continuum. 

%%%%%%%%%%%%%%%%%%%%%%%%%%%%%%%%%%%%%%%%%%%%%%%%%%%%%%%%%%%%%%%%%%%%%%%%%%%%%%

\subsection {Phenomenological window}
The phenomenological side of the correlation function is determined in a few steps: First, it is saturated by a full set of hadronic states carrying the same numbers as the interpolating current, and then the contribution of the ground state is isolated. As a result, we get 
\begin{equation} \label{had1}
\Pi^{Phe}_{\mu\nu}(p)=- \frac{\langle\psi_0|J_{\mu}|Z_c(p)\rangle \langle Z_c(p)|J^{\dagger}_{\nu}|\psi_0\rangle}{p^{*2}-m_{Z_c}^{*2}} + ... ,
\end{equation}
where $p^{*}$ is the in-medium momentum and  dots denote contributions arising from higher resonances and  continuum states. The decay constant or current-meson coupling is expressed  in terms of the polarization vector $\varepsilon_{\mu}$ of $Z_c$ as
\begin{equation} \label{had2}
\langle\psi_0|J_{\mu} |Z_c(p)\rangle = f^{*}_{Z_c} m_{Z_c}^{*} \varepsilon_{\mu},
\end{equation}
which further simplifies Eq. (\ref{had1}). By  summing over the  polarization vectors,  we can recast the phenomenological side of the correlation function into the form
\begin{equation}
\Pi_{\mu\nu}^{Phe}(p)=-\frac{m^{*2}_{Z_c} f^{*2}_{Z_c}}{p^{*2}-m^{*2}_{Z_c}} \Big[-g_{\mu\nu} +\frac{p_{\mu}^{*} p^{*}_{\nu} }{m^{*2}_{Z_c}} \Big] + ... .
\end{equation}
To proceed, we introduce two self energies: the scalar self-energy $\Sigma_s=m^{*}_{Z_c}-m_{Z_c}$ and   the vector self-energy $\Sigma_{\upsilon}$ appears in the expression of the in-medium momentum, $p_{\mu}^*=p_{\mu}-\Sigma_{\upsilon}u_{\mu}$ 
\cite{PhysRevC.46.1507}, where $  u_{\mu}$ is the four velocity vector of the cold nuclear medium. We work in the rest frame of the medium, i.e. $u_{\mu}=(1,0)$ .  As a result, one can write
\begin{eqnarray} \label{PiPhe}
\Pi_{\mu\nu}^{Phe}(p)=&-&\frac{ f^{*2}_{Z_c}}{p^2-\mu^2} \Big[-g_{\mu\nu} m^{*2}_{Z_c} +p_{\mu}p_{\nu} \nonumber \\
&-& \Sigma_{\upsilon}p_{\mu}u_{\nu} -\Sigma_{\upsilon}p_{\nu}u_{\mu}+\Sigma_{\upsilon}^2 u_{\mu}u_{\nu} \Big] + ...,\nonumber\\
\end{eqnarray}
where $\mu^2=m^{*2}_{Z_c}-\Sigma^2_{\upsilon} +2p_0\Sigma_{\upsilon}$, with $ p_0=p.u $ being the energy of the quasi-particle. The Borel transformed form (Borel with respect to $p^2$) of the phenomenological representation reads
\begin{eqnarray} \label{PiPhe2}
\Pi_{\mu\nu}^{Phe}(p)&=&f^{*2}_{Z_c}e^{-\mu^2/M^2} \Big[-g_{\mu\nu} m^{*2}_{Z_c} \nonumber \\
&+& p_{\mu}p_{\nu} - \Sigma_{\upsilon}p_{\mu}u_{\nu} -\Sigma_{\upsilon}p_{\nu}u_{\mu}+\Sigma_{\upsilon}^2 u_{\mu}u_{\nu} \Big]  \nonumber \\&+& ...,
\end{eqnarray}
where $M^2$ is the Borel mass parameter to be fixed later using the standard recipe of the method.

%%%%%%%%%%%%%%%%%%%%%%%%%%%%%%%%%%%%%%%%%%%%%%%%%%%%%%%%%%%%%%%%%%%%%%%%%%%%%%

\subsection {  OPE or QCD window}
The QCD side of the calculations can be obtained by inserting the interpolating current $J_{\mu}(x)$ into the correlation function and performing all possible contractions of the quark pairs. The resultant equation is a expression in terms of the in-medium light  quark  ($S_q^{ij}$) and heavy quark ($S_Q^{ij}$) propagators:
\begin{eqnarray}\label{qcd}
&&\Pi_{\mu\nu}^{QCD}(p)= -\frac{i}{2}\varepsilon_{abc} \varepsilon_{a'b'c'} \varepsilon_{dec} \varepsilon_{d'e'c'} \nonumber \\
&\times& \int d^4 x e^{ipx}\Big\{Tr\Big[\gamma_5 \tilde{S}_u^{aa'} (x)\gamma_5 S_c^{bb'}(x)\Big] \nonumber\\
&\times& Tr\Big[\gamma_{\mu} \tilde{S}_c^{e'e} (-x)\gamma_{\nu} S_d^{d'd}(-x)\Big] - Tr\Big[[\gamma_{\mu} \tilde{S}_c^{e'e} (-x) \nonumber\\
&\times& \gamma_5 S_d^{d'd}(-x)\Big]Tr\Big[\gamma_{\nu} \tilde{S}_u^{aa'} (x)\gamma_5 S_c^{bb'}(x)\Big] \nonumber \\
&-& Tr\Big[\gamma_5 \tilde{S}_u^{aa'} (x)\gamma_{\mu} S_c^{bb'}(x)\Big] Tr\Big[\gamma_5 \tilde{S}_c^{e'e} (-x)\gamma_{\nu} S_d^{d'd}(-x)\Big] \nonumber \\
&+& Tr\Big[\gamma_{\nu}  \tilde{S}_u^{aa'} (x)\gamma_{\mu} S_c^{bb'}(x)\Big] \nonumber \\
&\times&Tr\Big[\gamma_5 \tilde{S}_c^{e'e} (-x)\gamma_5 S_d^{d'd}(-x)\Big]  \Big\}_{|\psi_0\rangle},\nonumber \\
\end{eqnarray}
where $\tilde{S}_{q(c)}= C S_{q(c)}^{T}C$.

The in-medium light and heavy quark propagators in coordinate space in the fixed point gauge and $m_q\rightarrow 0$ limit are given  as
\begin{eqnarray}\label{ne1}
S_q^{ij}(x)&=&
\frac{i}{2\pi^2}\delta^{ij}\frac{1}{(x^2)^2}\not\!x
 + \chi^i_q(x)\bar{\chi}^j_q(0) \nonumber \\
&-&\frac{ig_s}{32\pi^2}F_{\mu\nu}^{ij}(0)\frac{1}{x^2}[\not\!x\sigma^{\mu\nu}+\sigma^{\mu\nu}\not\!x] +\cdots \, ,\\
\mbox{and}\nonumber\\
S_c^{ij}(x)&=&\frac{i}{(2\pi)^4}\int d^4k e^{-ik \cdot x} \left\{\frac{\delta_{ij}}{\!\not\!{k}-m_c}\right.\nonumber\\
&&\left.-\frac{g_sF_{\mu\nu}^{ij}(0)}{4}\frac{\sigma_{\mu\nu}(\!\not\!{k}+m_c)+(\!\not\!{k}+m_c)
\sigma_{\mu\nu}}{(k^2-m_c^2)^2}\right.\nonumber\\
&&\left.+\frac{\pi^2}{3} \Big\langle \frac{\alpha_sGG}{\pi}\Big\rangle\delta_{ij}m_c \frac{k^2+m_c\!\not\!{k}}{(k^2-m_c^2)^4}+\cdots\right\} \, . \nonumber \\
\end{eqnarray}
In the above equations, $\chi^i_q$ and $\bar{\chi}^j_q$ are the Grassmann background quark fields and
\begin{equation}
\label{ }
F_{\mu\nu}^{ij}=F_{\mu\nu}^{A}t^{ij,A}, ~~~~~A=1,2, ...,8,
\end{equation}
where $F_{\mu\nu}^A$ are classical background gluon fields, and $t^{ij,A}=\frac{\lambda ^{ij,A}}{2}$, with $
\lambda ^{ij, A}$ being  the standard Gell-Mann matrices. The next step is to use the expressions of the quark propagators in Eq. (\ref{qcd}). This leads to two main contributions: perturbative and nonperturbative. The perturbative contributions, which represent the short distance effects are calculated directly by applying the Fourier, Borel and continuum subtraction procedures. The nonperturbative or long distance contributions are expressed in terms of the in-medium quark, gluon and mixed condensates.  For the explicit expressions  of the condensates and the details of their expansions in terms of different operators see, for instance, Ref. \cite{AZIZI2018151}. After insertion of the in-medium condensates in terms of various operators, the Fourier and Borel transformations as well as continuum subtraction are also applied to the nonperturbative part of the QCD side.

The QCD side of the correlation function can be decomposed over the selected Lorentz structures as
\begin{eqnarray}\label{QCDcof}
\Pi_{\mu\nu}^{QCD}(p)&=&-\Upsilon^{QCD}_1 (p^2) g_{\mu\nu}  + \Upsilon^{QCD}_2 (p^2) p_{\mu}p_{\nu}    \nonumber \\
&-&  \Upsilon^{QCD}_3 (p^2) p_{\mu}u_{\nu}  -\Upsilon^{QCD}_4 (p^2) p_{\nu}u_{\mu}  \nonumber \\
&+& \Upsilon^{QCD}_5 (p^2) u_{\mu}u_{\nu},
\end{eqnarray}
where the invariant functions $\Upsilon^{QCD}_i$ ($i=1,...,5$) in Eq. (\ref{QCDcof}) are given in terms of the  following dispersion integrals:
\begin{equation}\label{Upsil}
 \Upsilon^{QCD}_{i} (p^2) = \int_{4m_c^2}^{\infty} \frac{\rho_i^{QCD}(s)}{s-p^2}ds,
\end{equation}
where $\rho_i^{QCD}(s)$ are  the two-point spectral densities related to the  imaginary parts of the selected coefficients.  The main purpose of the QCD side of the calculations is to calculate these spectral densities. To this end and after insertion of the quark propagators into the correlation function we use the relation
\begin{eqnarray}
\label{ }
\frac{1}{(x^2)^m}&=&\int \frac{d^Dk }{(2\pi)^D}e^{-ik \cdot x}i(-1)^{m+1}2^{D-2m}\pi^{D/2} \nonumber \\
&\times& \frac{\Gamma[D/2-m]}{\Gamma[m]}\Big(-\frac{1}{k^2}\Big)^{D/2-m},
\end{eqnarray}
to bring $ x $ to the exponential. In this way, the resultant expressions contain four four-integrals: one over four-$ k $ coming from the above relation one initially existing  in the correlation function (integral over four-$ x $) and the last two over four-$ k_1 $ and four-$ k_2 $ coming from the heavy quark propagators.  The next step is to perform the four integral over $ x $, which gives a Dirac delta function. We use this function to perform integral over $ k_1 $.  The remaining two integrals over four-$ k_2 $ and four-$ k $ are performed using the Feynman parametrization and the formula
\begin{equation}
\label{ }
\int d^4 \ell\frac{(\ell^2)^m}{(\ell^2+\Delta)^n}=\frac{i\pi^2 (-1)^{m-n} \Gamma[m+2]\Gamma[n-m-2]}{\Gamma[2]\Gamma[n] (-\Delta)^{n-m-2}}.
\end{equation} 
Finally, by applying the relation
\begin{equation}
\label{ }
\Gamma\Big[\frac{D}{2}-n\Big]\Big(-\frac{1}{\Delta}\Big)^{D/2-n}=\frac{(-1)^{n-1}}{(n-2)!}(-\Delta)^{n-2}ln[-\Delta],
\end{equation} 
and expansion of $ln[-\Delta] $, we get the imaginary parts of the $ \Upsilon^{QCD}_{i}  $ functions.

We apply the  Borel transformation with respect to  the variable $p^2$ and perform subtraction procedure according to the standard prescriptions of the method. As a result, we get 
\begin{equation}\label{Upsil2}
\mathbf{ \Upsilon}^{QCD}_{i}(M^2, s_0^*)= \int_{4m_c^2}^{s^{*}_0} ds \rho^{QCD}_{i}(s)e^{-\frac{s}{M^2}},
\end{equation}
where $s_0^*$ is the in-medium continuum threshold parameter  separating  the contributions of the ground state $Z_c$ and higher resonances and continuum. It  will be fixed later. The spectral density  related to each structure is the sum of the spectral densities of the perturbative (pert), two-quark ($qq$), two-gluon ($gg$) and mixed quark-gluon ($qgq$) parts: 
\begin{eqnarray}\label{ }
\rho_i^{QCD} (s)&=&\rho_i^{pert}(s) + \rho_i^{qq} (s) + \rho_i^{gg} (s) + \rho_i^{qgq} (s).
\end{eqnarray}
As an example, for the $g_{\mu\nu}$ structure, these spectral densities are collected in the Appendix.

To obtain the QCD sum rules for the mass, current coupling constant and vector self energy of the  $Z_c$ state, the coefficients of the  same structures  from both the $\Pi_i^{Phe}$ and $\Pi_i^{QCD}$ functions  are equated. We get the following sum rules for the physical quantities under consideration:
\begin{eqnarray}\label{SumR}
-m^{*2}_{Z_c}f^{*2}_{Z_c} e^{-\frac{\mu^2}{M^2}}& = & \mathbf{ \Upsilon}^{QCD}_{1}(M^2, s_0^*), \nonumber  \\
f^{*2}_{Z_c} e^{-\frac{\mu^2}{M^2}}& = & \mathbf{ \Upsilon}^{QCD}_{2}(M^2, s_0^*), \nonumber  \\
-\Sigma_{\upsilon}f^{*2}_{Z_c} e^{-\frac{\mu^2}{M^2}}& = &\mathbf{  \Upsilon}^{QCD}_{3}(M^2, s_0^*), \nonumber  \\
-\Sigma_{\upsilon}f^{*2}_{Z_c} e^{-\frac{\mu^2}{M^2}}& = & \mathbf{ \Upsilon}^{QCD}_{4}(M^2, s_0^*), \nonumber  \\
\Sigma^2_{\upsilon}f^{*2}_{Z_c} e^{-\frac{\mu^2}{M^2}}& = & \mathbf{ \Upsilon}^{QCD}_{5}(M^2, s_0^*).
\end{eqnarray}

%%%%%%%%%%%%%%%%%%%%%%%%%%%%%%%%%%%%%%%%%%%%%%%%%%%%%%%%%%%%%%%%%%%%%%%%%%%%%%

\section{Numerical Results}
In this section, the QCD sum rules  presented in Eq. (\ref{SumR})  are used to investigate the behavior of the  mass, current-meson coupling and vector self energy of $Z_c$  state in  cold nuclear matter.  To this end we need the values of  some input parameters and in-medium condensates entering the expressions of the obtained sum rules.  The  in-medium expectation values of different operators together with some other input parameters used in the calculations are: $\rho^{sat}=0.11^3 ~$GeV$^3$, $p_0=3887.2\pm2.3$ MeV \cite{PhysRevD.98.030001}, $ \langle q^{\dag} q\rangle_{\rho}=\frac{3}{2}\rho$, $ \langle \bar{q}q\rangle_{0}=(-0.241)^3$~GeV$^3$ \cite{IOFFE2006232},  $\langle \bar{q}q\rangle_{\rho}=\langle \bar{q}q\rangle_{0}+\frac{\sigma_{\pi N}}{2 m_q}\rho$ \cite{PhysRevC.47.2882}, $m_q=\frac{m_u+m_d}{2}=0.00345$~GeV \cite{PhysRevD.98.030001}, $\langle \frac{\alpha_s}{\pi}G^2\rangle_{0}=(0.33\pm0.04)^4~$GeV$^4$, $\langle \frac{\alpha_s}{\pi}G^2\rangle_{\rho}=\langle \frac{\alpha_s}{\pi}G^2\rangle_{0}-(0.65\pm 0.15)~$GeV$~\rho$, $\langle q^{\dag}iD_0 q\rangle_{\rho}=0.18~$GeV$~\rho$, $\langle \bar{q}iD_0q\rangle_{\rho}=0$, $\langle q^{\dag}iD_0 iD_0 q\rangle_{\rho_N}=0.031~GeV^2~\rho_N-\frac{1}{12}\langle q^{\dag}g_s\sigma G q\rangle_{\rho_N}$, $\langle \bar{q}g_s\sigma G q\rangle_{0}=m_0^2 \langle \bar{q}q\rangle_{0}$  \cite{COHEN1995221}, $m_0^2=0.8 ~$GeV$^2$ \cite{IOFFE2006232}, $\langle \bar{q}g_s\sigma G q\rangle_{\rho}=\langle \bar{q}g_s\sigma G q\rangle_{0}+ 3~GeV^2~\rho$,$\langle q^{\dag}g_s\sigma G q\rangle_{\rho}=-0.33 ~GeV^2~\rho$, $\langle q^{\dag}iD_0 iD_0 q\rangle_{\rho}=0.031~$GeV$^2~\rho-\frac{1}{12}\langle q^{\dag}g_s\sigma G q\rangle_{\rho}$ \cite{COHEN1995221,PhysRevC.47.2882}.  The pion-nucleon sigma term  $\sigma_{\pi N}=0.045$~GeV \cite{PhysRevD.87.074503} is used. The  quark masses are taken as  $m_u=2.16_{-0.26}^{+0.49}$~MeV, $m_d=4.67_{-0.17}^{+0.48}$~MeV and   $m_c= 1.27\pm 0.02$ GeV  \cite{PhysRevD.98.030001}.  

%And also, for the in-medium case  each condensate  is expanded up to the first order in nucleon density as $\langle\hat{O}\rangle_{\rho}=\langle\hat{O}
%%\rangle_{0} +\langle\hat{O}\rangle_{N}\rho$, where $\langle\hat{O}\rangle_{0}$ is the vacuum expectation value of the operator $  \hat{O}$ and $\langle\hat{O}%&\rangle_{N}$ is its expectation  value between one-nucleon states \cite{COHEN1995221,PhysRevC.47.2882}.

Besides these inputs, the sum rules in Eqs. (\ref{SumR}) require fixing of two auxiliary parameters: Borel mass parameter $M^2$ and the in-medium continuum threshold $s^*_0$. To proceed, we need to determine their working regions, such that the in-medium mass, current coupling the the vector self energy  show mild variations with respect to the changes in these parameters. The upper and lower limits for $M^2$  is determined via  imposing some conditions according to the standard prescriptions of the method (for details in the case of doubly heavy baryons, see, for instance, Refs.  \cite{PhysRevD.99.074012,PhysRevD.100.074004}). The working window for $s^*_0$ is obtained such that the maximum possible pole contribution is obtained and the physical observables show weak dependence on the  auxiliary parameters. These requirements lead to the working windows for auxiliary parameters  displayed  in Table (\ref{tab1}). 
\begin{table}[ht!]
\centering
\begin{tabular}{ll}
\hline\hline
% after \\ : \hline or \cline{col1-col2} \cline{col3-col4} ...
  $M^2$   & [3 - 5] GeV$^2$  \\
  $s^*_0$& $[17.6 -19.4]$ GeV$^2$ \\
\hline\hline
\end{tabular}
\caption{Working windows for  $M^2$ and $s^*_0$.}\label{tab1}
\end{table}
\begin{figure}[h!]
\centering
\begin{tabular}{c}
\epsfig{file=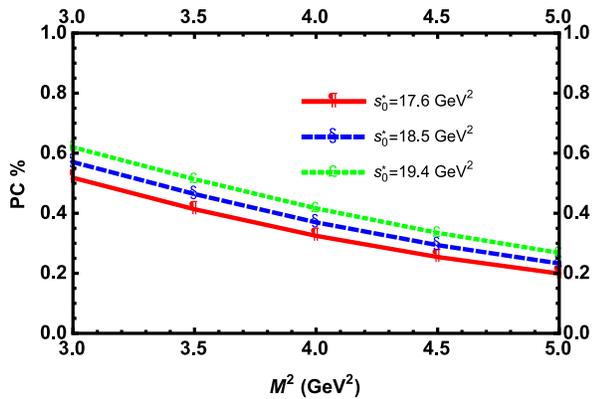,width=0.90\linewidth,clip=}  
\end{tabular}
\caption{The pole contribution in the $Z_c$ channel as a function of $M^2$ at saturation nuclear matter density and different fixed values of the in-medium continuum threshold.}\label{Fig1}
\end{figure}
As is seen from Fig. (\ref{Fig1}), the average pole contribution changes in the interval $[57\%-23\%]$ corresponding to $M^2=[3-5]$ GeV$^2$, which is reasonable in the case of tetraquarks. Our analyses show that the OPE  series of sum rules converge very nicely within the working windows of the auxiliary parameters.

Prior to the investigation of the behavior of the in-medium physical quantities, we would like to extract the vacuum mass value for $Z_c$ state.  The in medium sum rules in the limit $\rho\rightarrow 0$ lead to the average value of vacuum mass  as  $(3932^{+103}_{-84})$ MeV. We compare this value with the experimental data and other theoretical predictions in vacuum  in Table (\ref{tab2}). As is seen, our prediction for the vacuum mass of $Z_c$ obtained via in-medium sum rules in  $\rho\rightarrow 0$ limit  is consistent with other presented results within the errors arising mainly from the choice of the auxiliary parameters $M^2$ and $s_0^*$. Note that for Ref.  \cite{PILLONI2017200}, we presented only one of the results obtained via different scenarios. This result is the closest result to the average experimental value. 
\begin{table}[t]
{\begin{tabular}{lcl}\hline \hline 
&  Method &   $m_{Z_c}$  \\
\hline\hline
PS &  IMQCDSR ($\rho \rightarrow 0$) & $(3932^{+103}_{-84})$ MeV \\
\cite{PhysRevD.98.030001} & Experiment & $(3887.2\pm2.3)$ MeV \\
\cite{Cui_2014} &  QCDSR &  $(3.88\pm0.17)$ GeV \\
\cite{PhysRevD.89.054019} & QCDSR & $(3.91^{+0.11}_{-0.09})$ GeV \\
\cite{Wang2014} & QCDSR & $(3.89^{+0.09}_{-0.09} )$ GeV \\
\cite{PILLONI2017200} & AAD& $(3893.2^{+5.5}_{-7.7})$ MeV \\
\cite{PhysRevD.96.034026} & QCDSR & $(3901^{+125}_{-148})$ MeV \\
\cite{PhysRevD.87.116004} & QCDSR & $(3.86\pm0.27)$ GeV \\
\hline\hline
\end{tabular}}
\caption{Experimental result and theoretical predictions of different methods for the vacuum mass of $Z_c$ state. PS means present study, AAD means amplitude analysis of the data  and IMQCDSR refers to the in-medium QCD sum rules.}\label{tab2}
\end{table}

\begin{figure}[h!]
\centering
\begin{tabular}{cc}
\epsfig{file=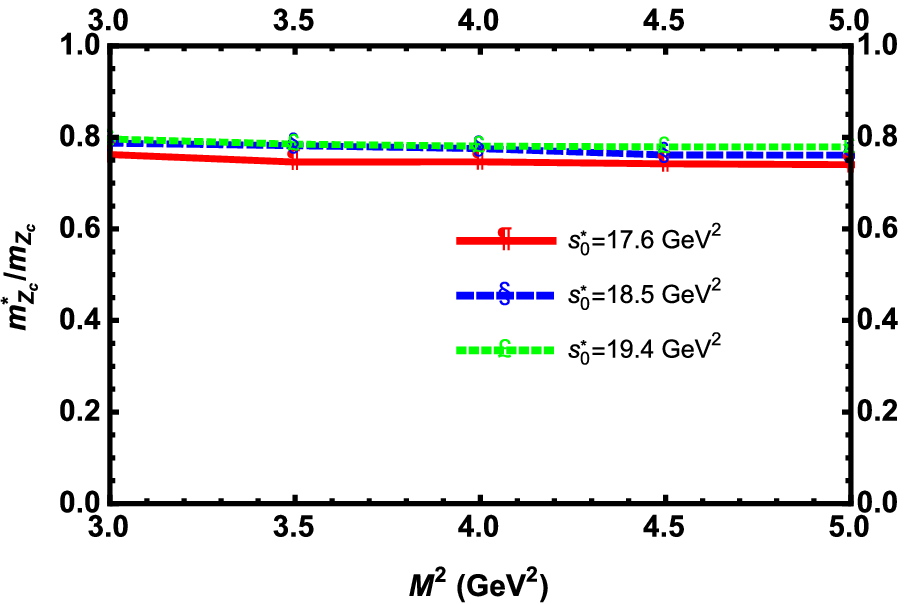,width=0.90\linewidth,clip=} \\
\epsfig{file=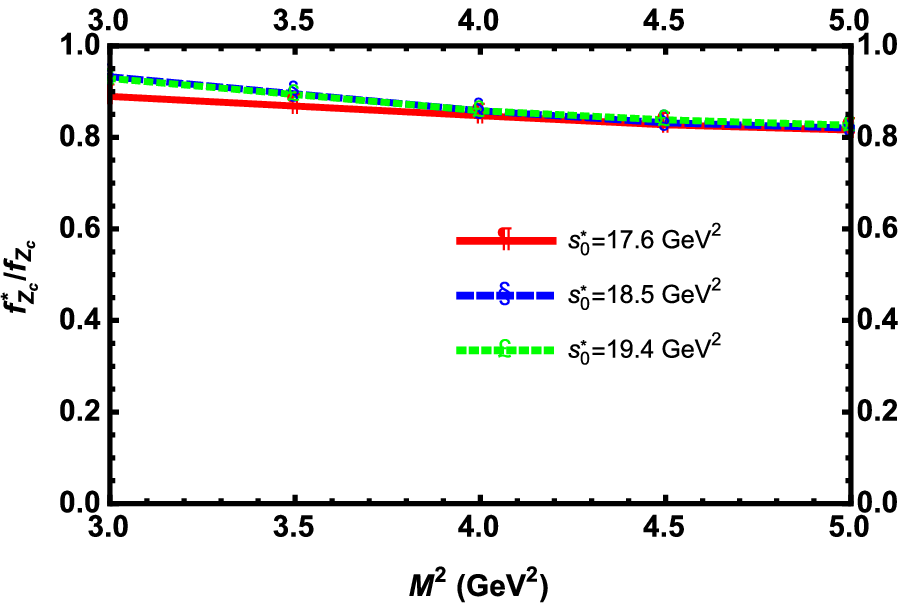,width=0.90\linewidth,clip=}  
\end{tabular}
\caption{ The ratios $m_{Z_c}^{*}/m_{Z_c}$ and   $f_{Z_c}^{*}/f_{Z_c}$ as  functions of $M^2$ at the saturated nuclear matter density, $\rho^{sat}=0.11^3$ GeV$^3$ and at fixed values of the continuum threshold.}\label{Fig2}
\end{figure}

\begin{figure}[h!]
\label{fig1}
\centering
\begin{tabular}{c}
\epsfig{file=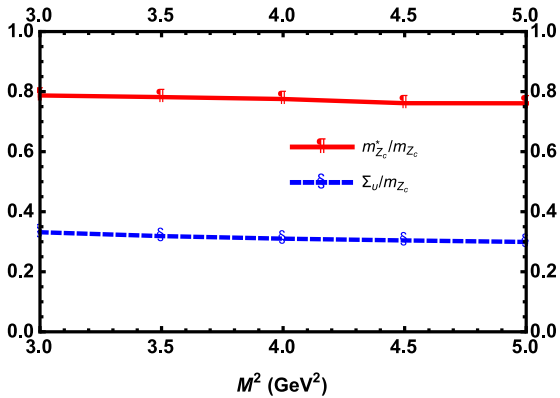,width=0.90\linewidth,clip=}  
\end{tabular}
\caption{The ratios $m_{Z_c}^{*}/m_{Z_c}$ and $\Sigma_{\nu}/m_{Z_c}$ as  functions of $M^2$ at average value of the continuum threshold and  at saturation nuclear matter density.}\label{Fig3}
\end{figure}

Now, we proceed to display the behavior of physical quantities  under consideration,  with respect to the Borel mass and continuum threshold parameters at saturation nuclear matter density, $\rho^{sat}=0.11^3$ GeV$^3$. In this respect, we plot the ratio of  the in-medium mass to vacuum mass, $m_{Z_c}^*/m_{Z_c}$,   and the ratio of the  in-medium current meson coupling to its vacuum value, $f_{Z_c}^*/f_{Z_c}$, in Fig. (\ref{Fig2}). As it is clearly seen, the  physical quantities show elegant stability against the changes in the parameters $M^2$ and $s_0^*$ in their  working regions. At the saturation nuclear matter density  the in-medium mass of $Z_c$ state decreases  to approximately  $77\%$ of its vacuum mass and  the shift in average current coupling value of the state due to nuclear medium is about $15\%$.  
In Fig. (\ref{Fig3}), the quantities  $m_{Z_c}^{*}/m_{Z_c}$ and $\Sigma_{\nu}/m_{Z_c}$ with respect to  $M^2$ at the average value of  continuum threshold and  saturation nuclear matter density are shown. As already mentioned, the average negative shift in the modified mass of $Z_c$ state due to cold nuclear matter (scalar self energy)  is about $23\%$ of the vacuum mass, whereas the vector self-energy, $\Sigma_{\nu}$, is obtained to be approximately $32\%$ of the vacuum value.

\begin{figure}[h!]
\centering
\begin{tabular}{cc}
\epsfig{file=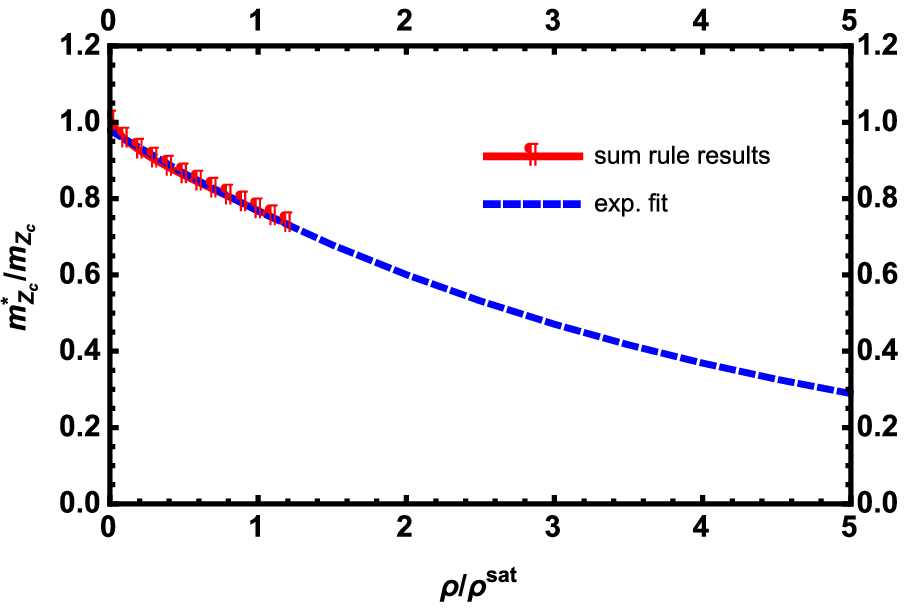,width=0.90\linewidth,clip=} \\
\epsfig{file=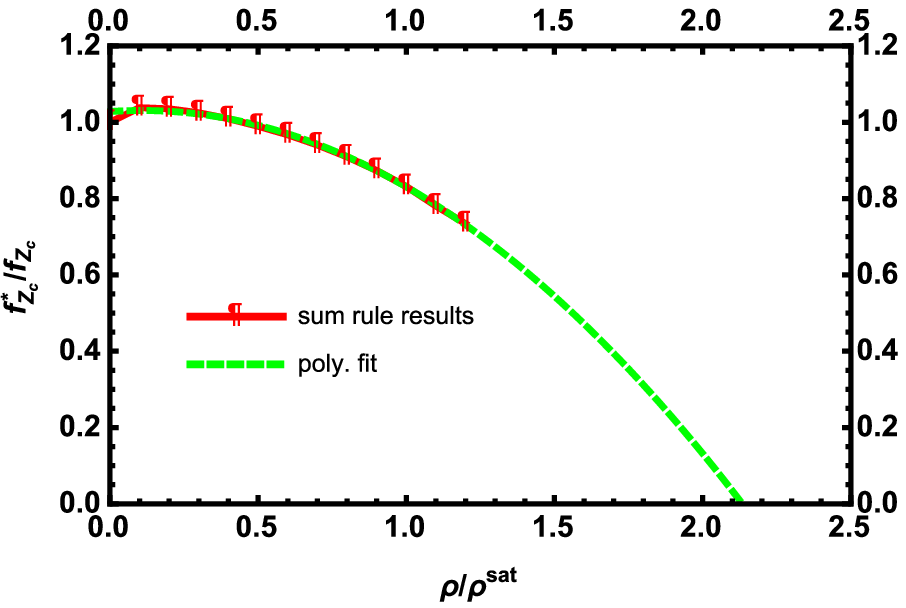,width=0.90\linewidth,clip=}  
\end{tabular}
\caption{The ratios $m_{Z_c}^{*}/m_{Z_c}$ and $f_{Z_c}^{*}/f_{Z_c}$ as  functions of $\rho/\rho^{sat}$  at mean values of the continuum threshold and  Borel parameter.}\label{Fig4}
\end{figure}

The main objective in  the present study is to investigate the variations of the physical quantities with respect to density of  the nuclear medium. To this end, we plot Fig. (\ref{Fig4}),  displaying $m_{Z_c}^*/m_{Z_c}$ and $f_{Z_c}^*/f_{Z_c}$  as  functions of $\rho/\rho^{sat}$ at  average values of the continuum threshold and  Borel parameter. The saturation nuclear matter mass density is $\rho^{sat} = 2.7 \times 10^{14}$ g/cm$^3$, which is equivalent to $\rho^{sat}=0.16$ fm$^{-3}$. However, we need to know the behavior of hadrons at higher densities, which will be accessible in heavy ion collision experiments. Neutron stars as natural laboratories  are very compact and dense that may produce hyperons and even heavy baryons and exotic states based on the processes that may occur inside them.  For a neutron star with mass $\sim 1.5 M_{\odot}$ the relevant core density is approximately ($2\rho^{sat}-3\rho^{sat}$) and for the mass $\sim 2 M_{\odot}$ the same density is about  $5\rho^{sat}$ \cite{Kim:2016yem}.
Therefore, we would like to discuss the behavior of the mass and coupling constants up to densities comparable with the densities of the neutron stars. However, as it is clear from  Fig. (\ref{Fig4}), the in-medium sum rules give reliable  results up to $\rho/\rho^{sat}=1$ and  $\rho/\rho^{sat}=1.1$ for $m_{Z_c}^{*}/m_{Z_c}$ and $f_{Z_c}^{*}/f_{Z_c}$, respectively. Hence, we need to extrapolate the results to include the higher densities. Our analyses show that the following fit functions 
well describe the ratios under consideration when the central values of the auxiliary and other input parameters are used:
\begin{eqnarray}\label{fitexp}
m_{Z_c}^{*}/m_{Z_c}=e^{-0.252 x},
\end{eqnarray}
and 
\begin{eqnarray}\label{fitpoly}
f_{Z_c}^{*}/f_{Z_c}=-0.251x^2+ 0.054x +1.028,
\end{eqnarray}
where $x= \rho/\rho^{sat}$. From Fig. (\ref{Fig4}), we see that the fit results coincide with the sum rules predictions at lower densities. The results show that the mass exponentially decreases with respect to $x$ and  reaches to roughly $30\%$ of the vacuum mass at a density $5\rho^{sat}$.  The coupling constant, however, rapidly changes with respect to $x$ and goes to zero at  $x=2.1$. This point may be considered as a pseudocritical density, at which hadrons are melted. 

\begin{figure}[h!]
\centering
\begin{tabular}{cc}
\epsfig{file=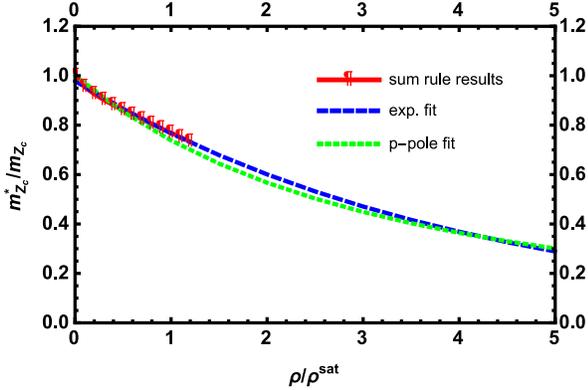,width=0.90\linewidth,clip=}   
\end{tabular}
\caption{Comparison between exponential and p-pole fit functions for $m_{Z_c}^{*}/m_{Z_c}$ changes with respect to $\rho/\rho^{sat}$.}\label{Fig7}
\end{figure}

\begin{figure}[h!]
\centering
\begin{tabular}{cc}
\epsfig{file=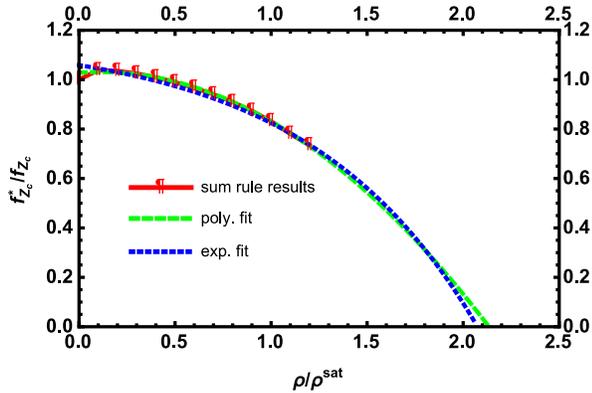,width=0.90\linewidth,clip=}   
\end{tabular}
\caption{Comparison between polynomial and exponential  fit functions for  $f_{Z_c}^{*}/f_{Z_c}$ changes with respect to $\rho/\rho^{sat}$.}\label{Fig8}
\end{figure}

It is necessary to check the results at higher densities using different fit functions that may end up in different predictions. Among various fit functions, the  
best ones are as follows:  are obtained to be the above fit functions: exponential fit for the mass ratio and polynomial fit for the coupling constant ratio. Our analyses show that another  set of  fit functions, although not as good as the above selected functions, describe the  behaviors of the quantities under study with respect to density, as well. These are the p-pole fit   for $m_{Z_c}^{*}/m_{Z_c}$:  
\begin{equation}
\frac{m_{Z_c}^{*}}{m_{Z_c}}= \frac{1}{\Big[1+\frac{x^2}{p ~a}\Big]^p},
\end{equation}
where $p=2.035$ and $a=3.056$,  and  exponential fit for $f_{Z_c}^{*}/f_{Z_c}$:
\begin{eqnarray}
f_{Z_c}^{*}/f_{Z_c}=1.169 -0.111 e^{- x/0.879}.
\end{eqnarray}
In  Figs. (\ref{Fig7}) and  (\ref{Fig8}), we compare the predictions of these fit functions with the predictions of the previous fit functions. In the case of mass ratio, although both fitting results show a similar behavior, the p-pole fitting leads to a result, which is $6.4\%$ larger than the exponential fitting result at the density $5\rho^{sat}$. In the case of the coupling constant, $f_{Z_c}^{*}/f_{Z_c}$ becomes zero at $\rho/\rho^{sat}= 2.13 $ and $ \rho/\rho^{sat}=2.07 $, for the polynomial and exponential fittings, respectively. The small differences in the predictions of different fit functions remain inside the uncertainties allowed by the method used.

\begin{figure}[h!]
\centering
\begin{tabular}{cc}
\epsfig{file=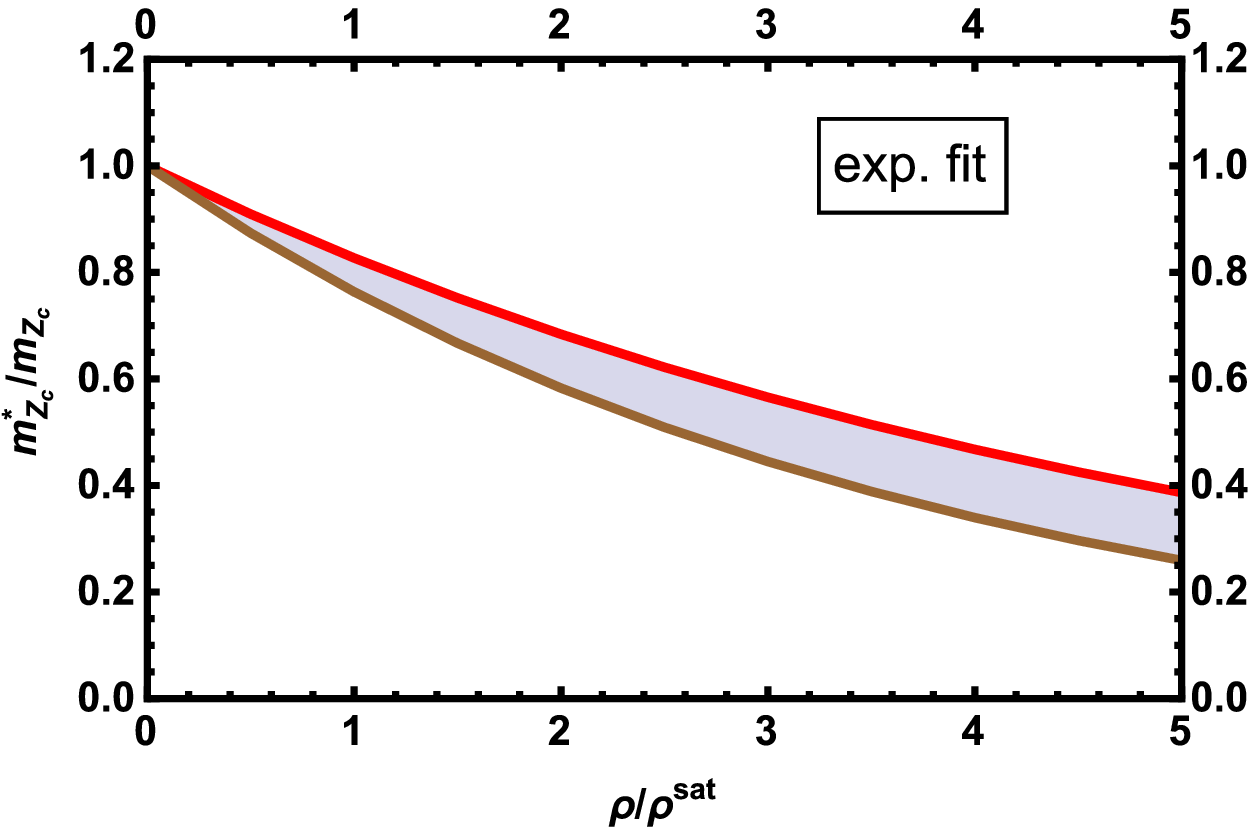,width=0.90\linewidth,clip=} \\
\epsfig{file=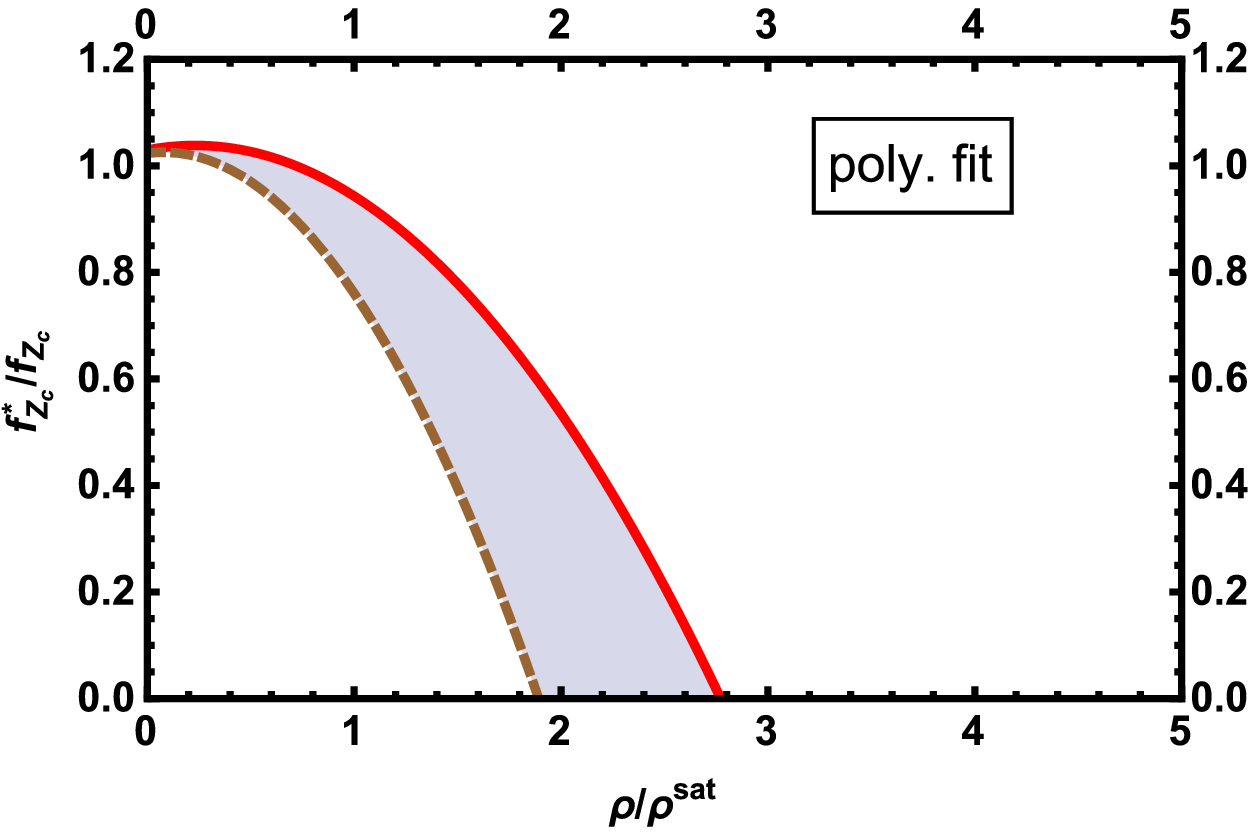,width=0.90\linewidth,clip=}  
\end{tabular}
\caption{The ratios $m_{Z_c}^{*}/m_{Z_c}$ and $f_{Z_c}^{*}/f_{Z_c}$ as  functions of $\rho/\rho^{sat}$  when the uncertainties of the auxiliary and other input parameters are considered.}\label{Fig9}
\end{figure}

Now, we add the uncertainties in the values of the auxiliary parameters and errors of other inputs to the predictions at higher densities. For the case of best fits for the mass and coupling ratios, i.e., Eqs. (\ref{fitexp}) and (\ref{fitpoly}) the swept areas presented in figure \ref{Fig9} are obtained. As it is seen, the values of the ratio $m_{Z_c}^{*}/m_{Z_c}$ vary in the region $ [0.26, 0.39] $ at  $ \rho/\rho^{sat}=5 $. The band of coupling ratio $f_{Z_c}^{*}/f_{Z_c}$ becomes zero and crosses the  $\rho/\rho^{sat}$ axis in the interval $ [1.9,2.8] $.

%%%%%%%%%%%%%%%%%%%%%%%%%%%%%%%%%%%%%%%%%%%%%%%%%%%%%%%%%%%%%%%%%%%%%%%%%%%%%%

\section{Discussion and  Conclusion}
Despite a lot of experimental and theoretical effort, the structure, quark-gluon organization and nature of most exotic states remain unclear. The tetraquark state $Z_c(3900)$ is among the charmonium-like resonances that deserve more investigations  in vacuum and a dense/hot medium in order to fix its nature. We considered it as a compact tetraquark and assigned a diquark-antidiquark structure to it with quantum numbers  $J^{PC}=1^{+-}$. We constructed in-medium sum rules to calculate its mass, current-coupling and vector self-energy in the cold nuclear matter. The result obtained for its mass at $\rho\rightarrow 0$ limit is compatible with the experimental data and vacuum theory predictions. At saturation density, the mass and current coupling receive negative shifts with respect to the vacuum values. These shifts amount to $23\%$ and $15\%$ for the mass and current coupling, respectively. This state receives  a repulsive vector self-energy with an amount of $32\%$ of its vacuum mass value at saturation nuclear matter density. 

The in-medium experiments aim to reach  higher densities. The neutron stars as the compact and dense objects are natural laboratories with densities $ 2-5 $ times greater than the saturation nuclear matter density. The processes that occur at high densities may produce different kinds of hadrons, even the exotic states. Such experiments that take place in high densities will serve a good opportunity to study the exotic states like $Z_c$. Hence, to provide some phenomenological predictions, we investigated  the behavior of the state under consideration with respect to density of the medium at higher densities. The in-medium sum rules are truncated at  around $\rho/\rho^{sat}=(1-1.1)$. Thus, we used some fit functions in order to extrapolate the results to a density around $5\rho^{sat}$. Our analyses show that the mass and current coupling constant exhibit non-linear behavior and decrease  with respect to density. The mass reaches to roughly $30\%$ of the vacuum mass at a density $5\rho^{sat}$ if the central values of auxiliary and other input parameters as well as the best fit are considered. When the uncertainties of the parameters are taken into account this value lies in the interval $ [26\%, 39\%]$. The current coupling however, rapidly decreases and goes to zero at a density $\approx 2.1\rho^{sat}$ if the central values of all parameters are considered, which may be considered as a pseudocritical density for melting of the exotic charmonium-like states. When the uncertainties in the values of the auxiliary and input parameters are taken into account, the band of $f_{Z_c}^{*}/f_{Z_c}$ becomes zero and crosses the  $\rho/\rho^{sat}$ axis in the region $ [1.9,2.8] $.

Investigation of properties of hadrons at finite temperature and density constitutes one of the main directions of the research in high energy and nuclear physics. Understanding the hadronic behavior under extreme conditions can help us not only understand the internal structure and nature of hadrons and gain knowledge of the QCD as the theory of the strong interaction, but also analyze the results of future experiments as well as understand different possible phases of matter. 

%%%%%%%%%%%%%%%%%%%%%%%%%%%%%%%%%%%%%%%%%%%%%%%%%%%%%%%%%%%%%%%%%%%%%%%%%%%%%%
\section*{ACKNOWLEDGMENTS}

The authors thank  TUBITAK for the  partial support provided under the Grant No. 119F094.
 
 %%%%%%%%%%%%%%%%%%%%%%%%%%%%%%%%%%%%%%%%%%%%%%%%%%%%%%%%%%%%%%%%%%%%%%%%%%%%%%
 %%%%%%%%%%%%%%%%%%%%%%%%%%%%%%%%%%%%%%%%%%%%%%%%%%%%%%%%%%%%%%%%%%%%%%%%%%%%%%

 \begin{widetext}
  \section{Appendix: Spectral Densities}
We present the explicit forms of the spectral densities for the $g_{\mu\nu} $ structure used in the calculations:
\begin{eqnarray}
\rho_1^{pert} (s)& = &-  \frac{1}{3072 \pi^6}\int_0^1 dz \int_0^{1-z} dw\frac{1}{\xi \vartheta^8} \Bigg[w z
 \Big[(s w z (w + z - 1) -  m_c^2 \Big[(w^3 + w^2 (2 z - 1)  \nonumber \\
&+&  2 w (z - 1) z  + (z - 1) z^2\Big]\Big]^2\Big[3 m_c^4 \Big[w^3 + w^2  
 (2 z - 1) + 2 w (z - 1) z + (z - 1) z^2\Big]^2 \nonumber \\ 
&-& 26 m_c^2 s w z \Big[w^4 + w^3 (3 z - 2) + w^2 (4 z^2
 -  5 z + 1)+w z (3 z^2 -  5 z + 2)+(z - 1)^2 z^2\Big]  \nonumber \\
&+& 35 s^2 w^2 z^2  (w + z-1)^2\Bigg] \Theta[L(s, z, w)], 
\end{eqnarray}
\begin{eqnarray}
\rho_1^{qq}(s) & = &\frac{1}{4\pi^4}\int_0^1 dz \int_0^{1-z} dw \Bigg\{ \frac{m_c w }{4\vartheta^5} \Bigg[3 m_c^4  \Big[w^3 
+ w^2 (2 z - 1) +  2 w (z - 1) z + (z - 1) z^2\Big]^2 \nonumber \\
&-& 10 m_c^2 s w z \Big[w^4 + w^3 (3 z - 2) + w^2 (4 z^2 - 5 z
+ 1) + w z (3 z^2 - 5 z + 2)  + (z - 1)^2 z^2\Big] \nonumber \\
&+& 7 s^2 w^2 z^2 (w + z - 1)^2\Bigg]\langle\bar{u}u\rangle_{\rho}
+  \frac{m_q m_c p_0 w^2 z \xi }{\vartheta^5} \Bigg[ 5 m_c^2 (w^3 + w^2 (2 z-1) \nonumber \\
&+& 2 w (z-1) z+(z-1) z^2) -7 s w z \xi \Bigg] 
 \langle u^{\dagger}u\rangle_{\rho} + \frac{m_c z }{4\vartheta^5} \Bigg[3 m_c^4  \Big[w^3 + w^2 (2 z - 1) \nonumber \\
&+&  2 w (z - 1) z + (z - 1) z^2\Big]^2 -10 m_c^2 s w z 
  \Big[w^4 + w^3 (3 z - 2) + w^2 (4 z^2 - 5 z +1)\nonumber \\
&+& w z (3 z^2 -  5 z + 2)+(z - 1)^2 z^2\Big]  
+ 7 s^2 w^2 z^2 (w + z - 1)^2\Bigg]\langle\bar{d}d\rangle_{\rho} \nonumber \\
&+&  \frac{m_q m_c p_0 w z^2 \xi }{\vartheta^5}  \Bigg[ 5 m_c^2 (w^3 + w^2 (2 z-1) 
+ 2 w (z-1) z+(z-1) z^2) -7 s w z \xi\Bigg] \nonumber \\
&\times& \langle d^{\dagger}d\rangle_{\rho} \Bigg\} \Theta[L(s, z, w)],
\end{eqnarray}
\begin{eqnarray}
\rho_1^{gg}(s) & = &\frac{1}{96\pi^4}\int_0^1 dz \int_0^{1-z} dw \Bigg\{ \Bigg[-i \frac{m_q m_c}{\vartheta^4} 
 \Big[m_c^2 \Big(4 w^6 z+ w^5 (24 z^2 - 8 z + 1) \nonumber \\
&+&  w^4 (52 z^3 - 44 z^2 +  6 z - 1) + w^3 z (60 z^3   
- 80 z^2 + 55 z - 2) + w^2 z^2 (40 z^3 - 68 z^2    \nonumber\\
&+&95 z -  34) + 6 w z^3 (2 z^3 - 4 z^2 + 13 z - 11)   
+ 33 (z - 1) z^4\Big) + w z \xi \Big(4 p_0^2 \big(4 w^3 z  \nonumber \\
&+&  w^2 (16 z^2- 4 z + 1) + 12 w (z - 1) z^2 + z^2\big) 
- 3 s \Big(4 w^3 z + w^2 (16 z^2 - 4 z + 1)  \nonumber\\
&+& 12 w (z - 1) z^2+ 17 z^2\Big)\Big) \Big]\Bigg]\langle u^{\dagger}iD_0u\rangle_{\rho} 
+ \frac{1}{1536(w-1)\xi^2 \vartheta^6}  \Bigg[m_c^4 (w-1) \Big(w^2 \nonumber\\ 
&+& w (z-1) + (z-1) z\Big)^2 \Big(24 w^8 z+12 w^7 
 (11 z^2-8 z+3) + 3 w^6 (112 z^3-156 z^2 \nonumber \\
&+&45 z-36)+2 w^5 (258 z^4-516 z^3+332 z^2 
-201 z+54)+ w^4 (516 z^5-1320 z^4+1574 z^3\nonumber\\
&-&1198 z^2+471 z-36)+w^3 z (336 z^5 
-1032 z^4+1707 z^3 -1769 z^2+858 z-132) \nonumber \\
&+& w^2 z^2 (132 z^5-468 z^4 + 1122 z^3-1253 z^2
+ 807 z-180)+4 w z^3 (6 z^5-24 z^4+152 z^3\nonumber \\
&-& 155 z^2+96 z-27)+8 z^4 (23 z^3 - 29 z^2 
+ 9 z-3)\Big)-m_q^2 s (w-1) w z \Big(24 w^{10} z \nonumber \\
&+&12 w^9 (13 z^2-12 z+3)+w^8 (492 z^3-852 z^2 
-89 z-180)+w^7 (984 z^4-2424 z^3+283 z^2 \nonumber \\
&+&672 z+360)+w^6 (1368 z^5-4296 z^4+2149 z^3 
+1121 z^2-270 z-360)+w^5 (1368 z^6\nonumber \\
&-&-5160 z^5 + 4057 z^4-420 z^3+275 z^2-676 z+180) 
+w^4 (984 z^7-4296 z^6+4151 z^5 \nonumber \\
&-&809 z^4+1804 z^3-1901 z^2+615 z-36)+w^3 z (492 z^7
-2424 z^6+2485 z^5+834 z^4\nonumber \\
&+&43 z^3 -2396 z^2 + 1098 z-132)+w^2 z^2 (156 z^7-852 z^6+898 z^5 \nonumber \\
&+&1887 z^4-2785 z^3-27 z^2+903 z-180)
+12 w (z-1)^2 z^3 (2 z^5-8 z^4+18 z^3+89 z^2 \nonumber \\
&-&12 z-9)+8 (z-1)^3 z^4 (23 z^2+30 z+3)\Big) 
-12 s w z^2 (w+z-1)^3 \Big(w^5 ((17 s+4) z-4) \nonumber \\
&+&w^4 (5 (3 s+2) z^2-2 (17 s+4) z+2)
+ w^3 z (s (32 z^2-13 z+17)+4 (3 z^2-4 z+1)) \nonumber \\
&+&2 w^2 z^2 (s (16 z^2-32 z-1)+7 z^2-10 z+3) 
- 8 w (z-1) z^3 (4 s-z+1) + 2 w^6 \nonumber \\
&+&4 (z-1)^2 z^4\Big)\Bigg]\Big\langle \frac{\alpha_s}{\pi}G^2\Big\rangle_{\rho}\Bigg\} \Theta[L(s, z, w)],  
\end{eqnarray}
and
\begin{eqnarray}
\rho_{1}^{qgq}(s)  &=&\int_0^1 dz \int_0^{1-z} dw  \Big[\rho_{1,1}^{qgq}(s) 
+\rho_{1,2}^{qgq}(s)  + \rho_{1,3}^{qgq}(s)  +\rho_{1,4}^{qgq}(s) 
+\rho_{1,5}^{qgq}(s)\Big]\Theta[L(s, z, w)],
\end{eqnarray}
where
\begin{eqnarray}
\rho_{1,1}^{qgq}(s) & = &\frac{1}{96\pi^4} \Bigg[-\frac{m_c}{\kappa^6} \Big[m_c^2 \Big(4 w^6 z + w^5 (24 z^2 - 8 z
+  1) + w^4 (52 z^3 - 44 z^2 - 2 z - 1) + w^3 z \nonumber \\
&\times& (60 z^3 - 80 z^2 + 31 z + 6) +  w^2 z^2 (40 z^3  
- 68 z^2+ 71 z - 18) + 2 w z^3 (6 z^3 - 12 z^2 \nonumber \\
&+& 35 z - 29) + 33 (z - 1) z^4\Big) \Big(w^2 + w (z - 1) 
+ (z - 1) z\Big)^3 + w z (w + z - 1) \Big(4 p_0^2 (4 w^9 z  \nonumber \\
&+& w^8 (28 z^2 - 16 z + 1) + 3 w^7 (28 z^3 - 36 z^2 
+ 25 z - 1)  +  w^6 (160 z^4 - 300 z^3 + 371 z^2  \nonumber \\
&-& 169 z + 3) +  w^5 (208 z^5 - 508 z^4 + 838 z^3 
- 694 z^2 + 157 z - 1) + w^4 z (192 z^5 - 564 z^4  \nonumber \\
&+& 1116 z^3 - 1276 z^2 + 567 z -  51) + 2 w^3 z^2  
 (62 z^5 - 210 z^4+ 463 z^3 - 642 z^2 + 409 z - 82) \nonumber \\
&+&  w^2 (z - 1)^2 z^3 (52 z^3 - 92 z^2 + 239 z  
- 164)+    3 w (z - 1)^3 z^4 (4 z^2 - 4 z + 17)  \nonumber \\
&+& (z - 1)^3 z^5) - s \big(12 w^9 z + w^8 (84 z^2 - 48 z 
+  3)+w^7 (252 z^3 - 324 z^2 + 121 z - 9) \nonumber \\
&+& w^6 (480 z^4 - 900 z^3 + 713 z^2 - 195 z + 9)
+ w^5 (624 z^5 - 1524 z^4 + 1730 z^3 - 986 z^2  \nonumber \\
&+& 159 z - 3) + w^4 z (576 z^5 - 1692 z^4 +2500 z^3  
- 2060 z^2+  709 z - 49) + 2 w^3 z^2 (186 z^5  \nonumber \\ 
&-& 630 z^4 + 1125 z^3 - 1218 z^2 + 635 z - 98)  
+ w^2 (z - 1)^2 z^3 (156 z^3 - 276 z^2 + 589 z - 292)  \nonumber \\
&+& w (z - 1)^3 z^4 (36 z^2 -36 z + 193)
+  51 (z - 1)^3 z^5\big)\Big)\Big] \Bigg]\langle \bar{u}iD_0iD_0u\rangle_{\rho},
\end{eqnarray} 
\begin{eqnarray}
\rho_{1,2}^{qgq}(s)  & = &\frac{1}{12\pi^4} \Bigg[\frac{m_c w z^2 \xi}{\vartheta^7} \Big[m_c^2 (w+z)\Big(w^2  
+w (z-1)+(z-1)z\Big)^3 -wz\xi \nonumber \\
&\times&\Big(8 p_0^2\big(3w^4+w^3(7z-6)+w^2(7z-6)
+w^2(10z^2-14z+3)+wz(6z^2-13z+7)\nonumber \\
&+&3(z-1)^2z^2\big)-s\big(5w^4+2w^3(6z-5)
+w^2(17z^2-24z+5)+2wz(5z^2-11z\nonumber \\
&+&6)+5(z-1)^2z^2\big)\Big)
\Big] \Bigg]\langle \bar{d}iD_0iD_0d\rangle_{\rho},
\end{eqnarray} 
\begin{eqnarray}
\rho_{1,3}^{qgq}(s)  &=&\frac{1}{96\pi^4}  \Bigg[ \frac{m_c}{4 \vartheta^7} \Big[ w z (w+z-1) 
\Big(2 p_0^2 (4 w^9 z+w^8 (28 z^2 -16 z+1) \nonumber \\
&+&3 w^7 (28 z^3-36 z^2+25 z-1) 
+ w^6 (160 z^4-300 z^3+371 z^2 \nonumber \\
&-& 169 m_cz+3)+w^5 (208 z^5-508 z^4 
+ 838 z^3-694 z^2+157 z-1) + w^4 z \nonumber \\
&\times&(192 z^5-564 z^4+1116 z^3-1276 z^2 
+567 z - 51)+2 w^3 z^2 (62 z^5-210 z^4 \nonumber \\
&+& 463 z^3-642 z^2 + 409 z-82)
+ w^2 (z-1)^2 z^3 (52 z^3-92 z^2+239 z \nonumber \\
&-& 164)+3 w (z-1)^3 z^4 (4 z^2-4 z+17) 
+ (z-1)^3 z^5) + s (4 w^9 z+w^8 (28 z^2 \nonumber \\ 
&-& 16 z-59)+w^7 (84 z^3-108 z^2 - 401 z 
+ 177)+w^6 (160 z^4-300 z^3 -1197 z^2 \nonumber \\
&+& 1259 z-177)+w^5 (208 z^5-508 z^4 
- 2022 z^3+3534 z^2 -1271 z+59) \nonumber \\
&+& w^4 z (192 z^5-564 z^4-2204 z^3 + 5368 z^2 
- 3185 z+425)+2 w^3 z^2 (62 z^5-210 z^4 \nonumber \\
&-& 751 z^3+2380 z^2-1945 z+464) 
+ w^2 (z-1)^2 z^3 (52 z^3-92 z^2-861 z+760) \nonumber \\ 
&+&w (z-1)^3 z^4(12 z^2-12 z-173) 
+ 25 (z-1)^3 z^5)\Big)-m_c^2 (w^2+w (z-1) \nonumber \\
&+&(z-1) z)^3 \Big(4 w^6 z+w^5 (24 z^2-8 z-35)
+ w^4 (52 z^3-44 z^2-170 z+35)+w^3 z (60 z^3 \nonumber \\
&-&80 z^2-353 z+174)+w^2 z^2 (40 z^3-68 z^2
-301 z  +234)+2 w z^3 (6 z^3-12 z^2-37 z \nonumber \\
&+& 43)+9 (z-1)z^4 \Big) \Big] \Bigg]\langle \bar{u}g_s\sigma Gu\rangle_{\rho},
\end{eqnarray}
\begin{eqnarray}
\rho_{1,4}^{qgq}(s)  &=&\frac{1}{48\pi^4}  \Bigg[ \frac{m_c w z^2 \xi}{\vartheta^7} \Big[13m_c^2 (w+z)\Big(w^2  
+w (z-1)+(z-1)z\Big)^3 +wz\xi\nonumber \\
&\times&\Big(4 p_0^2\big(3w^4+w^3(7z-6)+w^2(10z^2
-14z+3)+wz(6z^2-13z+7)\nonumber \\
&+&3(z-1)^2z^2\big)-s\big(31w^4+w^3(66z-62)
+w^2(97z^2-132z+31)+2wz(31z^2-64z\nonumber \\
&+&33)+31(z-1)^2z^2\big)\Big) \Big] \Bigg]\langle \bar{d}g_s\sigma Gd\rangle_{\rho},
\end{eqnarray}
and
\begin{eqnarray}
\rho_{1,5}^{qgq}(s)  &=&\frac{1}{192\pi^4}  \Bigg[ \frac{p_0z}{\vartheta^4} \Big[m_c^2\Big(16w^7 z+w^6(48z^2  
-48z+31)+w^5(80z^3-128z^2+137z -62)\nonumber \\
&+&w^4(80z^4-176z^3+237z^2   
-164z+31)+w^3z(48z+4-128z^3 + 223z^2-190z+59)\nonumber\\
&+&w^2z^2(16z^4 
-48z^3+109z^2-110z+33)+wz^3(25z^2-38z+13) \nonumber \\
&+&8(z-1)^2z^4\big) 
\Big] \Bigg]\langle u^{\dag}g_s\sigma Gu\rangle_{\rho}.
\end{eqnarray}
The shorthand notations  in the above equations are given as
\begin{eqnarray}
\xi&=& (w+z-1), \nonumber \\
\vartheta&=&(w^2+w (z-1)+(z-1) z), \nonumber \\
L[s, w,  z] &=& \frac{((-1 + w)(-(swz(-1 + w + z)) + 
       m_c^2 (w^3 + 2 w (-1 + z) z + (-1 + z) z^2 + 
          w^2 (-1 + 2z))))}{(w^2 + w (-1 + z) + (-1 + z) z)^2}.
\end{eqnarray}
\end{widetext}

  \begin{widetext}
\bibliography{refs.bib}
  \end{widetext}
%%%%%%%%%%%%%%%%%%%%%%%%%%%%%%%%%%%%%%%%%%%%%%%%%%%%%%%%%%%%%%%%%%%%%%%%%%%%%%

\end{document}